# Dynamic Heterogeneity and Facilitation in Sheared Granular Materials: Insights from 3D Triaxial Testing

Kwangmin Lee [a], Brett S. Kuwik [a], Ryan C. Hurley [a, b *]

[a] *Department of Mechanical Engineering, Johns Hopkins University, Baltimore, Maryland 21218, USA*
[b] *Hopkins Extreme Materials Institute, Johns Hopkins University, Baltimore, Maryland 21218, USA*

ABSTRACT: Strain localization in granular materials arises from complex microscale dynamics, including intermittent particle rearrangements and spatiotemporally correlated deformation. While dynamic heterogeneity (DH) and dynamic facilitation (DF) have been widely studied in two-dimensional amorphous materials, their prevalence in three-dimensional (3D) granular systems remains unclear. Here, we performed a 3D triaxial compression test with *in-situ* X-ray computed tomography to track particle-scale kinematics across small and large strain increments. We analyzed deviatoric strain, volumetric strain, and non-affine motion fields, computed four-point spatial dynamic correlation functions to probe DH, quantified DF through a facilitation ratio, and assessed temporal persistence of local dynamics using four-point temporal dynamic correlations. Across large strain increments, DH and DF emerge strongly in the transition regime between the initially elastic response and the critical state regime, but weaken or become statistically insignificant within the shear band at the critical state, indicating a qualitative change in microscale dynamics upon localization. In contrast, under small increments, both measures are suppressed across all regimes. These results demonstrate that correlated dynamics depend strongly on both strain increment and deformation regime. This work provides the first comprehensive investigation of DH and DF in 3D granular materials and highlights their strain-increment and regime-dependent behaviors, establishing a connection to glassy dynamics in amorphous solids.

KEYWORDS: Granular materials, Strain localization, Dynamic heterogeneity, Dynamic facilitation.

E-mail address: rhurley6@jhu.edu

## 1 INTRODUCTION

Strain localization in granular materials is commonly observed in geophysical failures such as landslides and earthquakes. To mitigate such failures, numerous studies have investigated the underlying mechanisms of strain localization. In continuum-based approaches, classical constitutive models such as the Mohr–Coulomb and Drucker–Prager models are often employed. However, these models assume homogeneity in the material—a simplification that breaks down once shear bands form after the elastic regime of deformation. Shear bands, characterized by localized deformation and particle rearrangements, introduce significant heterogeneity into the material response, thereby limiting the predictive capability of traditional models. To address this limitation, two-scale models have been developed to separately account for the mechanical behavior inside and outside shear bands [1]. Although this approach has improved modeling accuracy, the microscale behavior within shear bands—where most deformation is concentrated—remains poorly understood. This gap hinders a comprehensive understanding of granular mechanics, especially during triaxial compression.

A deeper understanding of such microscale behavior can be gained by drawing analogies with amorphous solids, where nonhomogeneous plasticity has been extensively studied. In amorphous solids, efforts to understand nonhomogeneous plastic behavior have focused on microscopic dynamics such as particle rearrangements. The concept of shear transformation zones, introduced by Falk and Langer [2], deals with localized regions of particle rearrangement, often quantified by the $D^2_{min}$ metric. On a statistical level, these rearrangements exhibit dynamic heterogeneity (DH)—a phenomenon characterized by spatially heterogeneous and temporally intermittent dynamics [3]. Experiments on colloidal glasses provided direct evidence of such heterogeneity, where clusters of fast-moving particles grow in prevalence and size near the glass transition [4,5]. Similar behavior has since been observed across diverse systems, from supercooled liquids to metallic glasses, with correlation lengths and relaxation times increasing sharply near the glass transition [6–8]. To account for the microscopic origin of DH in glassy systems, the dynamic facilitation (DF) framework has been proposed, which posits that local rearrangements facilitate subsequent rearrangements in neighboring particles, leading to spatiotemporal avalanches of activity [9,10].

In granular materials, similar signatures of DH and DF have been reported. Dauchot et al. [11] provided direct experimental evidence of DH in dense granular assemblies close to jamming, demonstrating finite dynamical correlation lengths through four-point correlation functions. Keys et al. [12] further showed that these dynamical lengths grow as the system approaches the jamming transition, in direct analogy with the glass transition. Le Bouil et al. [13] showed that early-stage strain localization emerges as transient microbands, distinct from the final persistent shear band, and Viallon-Galinier et al. [14] demonstrated that the observed localization patterns depend on the strain increment: very small increments yield isolated strain peaks, whereas larger increments lead to continuous shear bands. Candelier et al. [15,16] provided direct experimental evidence that cage-jump events cluster into avalanches which underlie collective relaxation. They showed that DF plays a central role in these avalanche processes [15], but also that its contribution becomes progressively weaker when approaching the granular glass transition [16]. Houdoux et al. [17] reported micro-slip triggering resembling aftershock sequences, which qualitatively aligns with DF-like behavior. Collectively, these studies demonstrate that granular media exhibit both DH and DF.

Explicit investigation of DH and DF in granular materials remains scarce under three-dimensional (3D) conditions. Prior work has largely been restricted to two-dimensional (2D) biaxial or simple shear setups, and it is still unclear how these phenomena evolve under triaxial compression. Although recent 3D X-ray computed tomography (XRCT) and discrete element method (DEM) studies have provided valuable insights into strain localization and the micromechanics of shear bands, including grain breakage [18–20], these works did not explicitly examine DH or DF. In particular, the development of DH and DF across the elastic, transition, and critical-state regimes, as well as their behavior inside and outside shear bands, has not been studied. This study aims to answer the following questions: (i) how do DH and DF vary with strain increment? (ii) how do DH and DF differ across deformation regimes? This work provides the first systematic 3D experimental investigation of DH and DF under triaxial compression, explicitly examining their strain-increment and regime-dependent behavior.

In this study, we conducted a triaxial compression test on granular materials and used *in-situ* 3D XRCT to obtain tomographic volumes at both large and small strain increments. We analyzed the microscale behavior using deviatoric strain, volumetric strain, and $D^2_{min}$ fields, and further examined DH through spatial four-point dynamic correlation functions, quantified DF using a facilitation ratio, and evaluated the temporal persistence of local dynamics. Through this approach, we bridge the gap between 2D and 3D studies and establish new connections between granular mechanics and glassy dynamics.

## 2 MATERIAL AND METHODS

### 2.1 Sample Preparation

A single-crystal alpha-quartz ($SiO_2$) block, hydrothermally grown and supplied by Sawyer Technical Materials, LLC, was used to prepare the sample. The block was crushed, and particles smaller than 300 $\mu$m were further processed in a custom air mill operating at 20 psi to round their edges, following the method described in Ref. [21]. After milling, the material was sieved to collect grains with diameters between 150 $\mu$m and 300 $\mu$m. Figure 1 summarizes the grain size and shape characteristics. In Figure 1A, the cumulative distribution of equivalent grain diameter is shown, where the equivalent diameter is calculated as $(6 \times N_p \times L_p^3/\pi)^{1/3}$, with $N_p$ representing the number of pixels in a grain and $L_p$ the pixel length. The median diameter was approximately 230 $\mu$m. Figure 1B presents the cumulative fraction of grains exceeding a given sphericity, calculated using Wadell's definition [22], $\pi^{1/3}(6V_p)^{2/3}/A_p$, where $V_p$ and $A_p$ denote the grain volume and surface area, respectively, measured using MATLAB's regionprops3 function. Most grains had sphericity values between 0.75 and 0.95, indicating that the air-milling process produced relatively smooth and rounded particles.

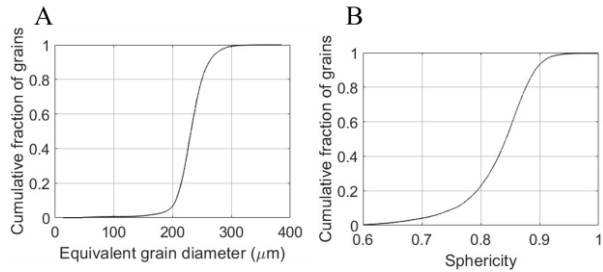

Figure 1. Morphology of the synthetic quartz particles used in this study: (A) cumulative distribution of equivalent grain diameter and (B) cumulative distribution of sphericity.

### 2.2 Experimental setup

A triaxial compression experiment was conducted using a custom-built High-Pressure TriAxial COmpression Apparatus (HP-TACO), illustrated in Fig. 2A. Full details of the device are available in Ref. [23], with prior applications described in Refs. [18,19]. For the test, the specimen was contained within a thin polymer cylinder (diameter 5 mm) positioned at the chamber base, into which the granular material was carefully poured. A top pin was then placed over the sample, the chamber was sealed, and a lateral confining pressure of 10 MPa was applied using water. Then, axial loading was applied under displacement control by slowly lowering an actuator-driven platen onto the top pin. At selected axial strain levels, loading was halted to carry out 3D XRCT scans using an RX Solutions EasyTom 150/160 micro-CT system (160 kV source). Examples of the sample configuration before and after deformation are provided in Fig. 2B.

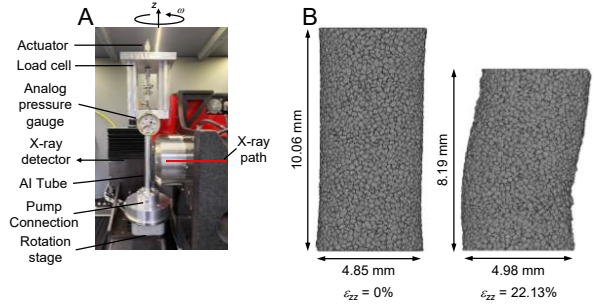

Figure 2. (A) High-Pressure TriAxial COmpression Apparatus (HP-TACO) used for triaxial testing. (B) 3D X-ray tomographic images of the granular sample before and after axial compression.

## 3 RESULTS

### 3.1 Macroscale behavior

The macroscale deviatoric stress–strain response from the experiment is presented in Figure 3. Deviatoric stress was calculated by taking the axial force measured with a load cell, dividing it by the specimen's initial top surface area, and subtracting the lateral confining pressure of 10 MPa, which remained nearly constant throughout the test. Axial strain was calculated as the displacement of the top platen divided by the specimen's initial height.

In most deformation steps, a relatively large axial strain increment of 2.45% was applied. To evaluate the influence of increment size, much smaller steps of 0.01% were used in steps 5–9 and 13–17. The stress–strain curve shows an almost linear relationship in the elastic regime (steps 1–3), a softening behavior in the transition regime (steps 3–9), and a plateau in the critical state regime (steps 9–19).

The stress values in Figure 3 were measured immediately after each loading step and do not include any post-loading stress relaxation. Although relaxation took place during the subsequent XRCT scans at all steps, it is not shown here in order to keep the stress–strain curve clear. The pronounced drop in stress from step 5 to step 6 occurred because the relaxation following the large strain increment at step 5 was much greater than the stress increase during the small strain increment between steps 5 and 6. During the subsequent small-increment steps (steps 6–9), the increase in stress was comparable to, or even smaller than, the magnitude of relaxation, leading to a continued reduction in measured stress. A similar trend was seen in steps 13–17, where small strain increments produced stress changes comparable to or smaller than the relaxation magnitude. For comparison, relaxation after large strain increments has also been reported in earlier studies using the same setup with different samples [18,19].

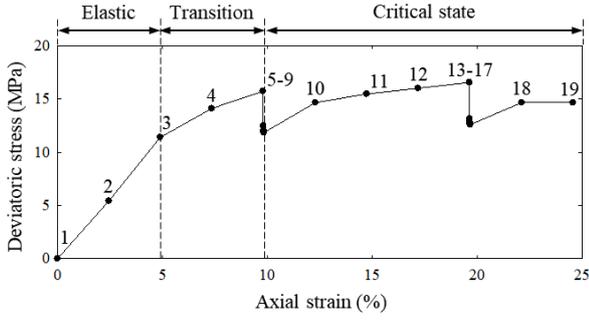

Figure 3. Macroscale deviatoric stress versus axial strain from triaxial compression. Numbers indicate deformation steps.

### 3.2 Microscale behavior

During the experiment, we analyzed microscale behavior by calculating incremental deviatoric strain, incremental volumetric strain, and non-affine motion fields between each pair of consecutive deformation steps (e.g., 1–2, 2–3, …, 18–19). These fields were derived from 3D tomographic volumes reconstructed in X-Act (version 23.04, RX Solutions) using a filtered backprojection algorithm with a Tukey filter (80% cutoff) and sinus apodization at 75%. The initial isotropic voxel size of 5.92 $\mu$m was reduced to 11.84 $\mu$m through twofold downsampling.

At every deformation step, reconstructed images were binarized with a fixed, heuristically chosen threshold. Grains were segmented via a distance transform followed by watershed segmentation, and small spurious fragments were removed using volume-based filtering in MATLAB. Grain-resolved digital volume correlation was then carried out in SPAM [24], treating each labeled grain as an independent correlation volume. This procedure yielded incremental displacement, rotation, and deformation gradient tensors for each grain.

From these kinematic data, we computed the incremental deviatoric strain ($\Delta\varepsilon_{dev}$) and incremental volumetric strain ($\Delta\varepsilon_{vol}$) fields. Local strain was calculated for each tetrahedron by performing 3D Delaunay triangulation on particle centers. For each tetrahedron, nodal displacement vectors (corresponding to particle-center displacements) were used to calculate the displacement gradient tensor, and subsequently the deformation gradient $\mathbf{F}$, following the formulation in Ref. [25]. The symmetric Lagrangian strain tensor was then obtained as

$$\mathbf{E} = \frac{1}{2}(\mathbf{F}^T \mathbf{F} - \mathbf{I}) \quad (1)$$

and its deviatoric component was calculated by subtracting the volumetric part:

$$\mathbf{E}_{dev} = \mathbf{E} - \frac{1}{3}\mathrm{tr}(\mathbf{E})\mathbf{I} \quad (2)$$

Tetrahedra with edge lengths greater than 1.5 mm were excluded to avoid artifacts from poorly shaped elements. The incremental deviatoric strain $\Delta\varepsilon_{dev}$ was defined as the Frobenius norm of $\mathbf{E}_{dev}$ and the incremental volumetric strain $\Delta\varepsilon_{vol}$ was defined as $\mathrm{tr}(\mathbf{E})$.

Figure 4 illustrates a cross-sectional view of the incremental deviatoric strain ($\Delta\varepsilon_{dev}$) field. In the early elastic regime (steps 1–2), strain is nearly homogeneous with no distinct clusters. However, in the late elastic regime (steps 2–3), faint heterogeneity in the form of clustered regions of high incremental deviatoric strain appears, implying that microscale DH can precede macroscopic softening, as also indicated in Sec. 3.3 by the high value of the dynamic correlation length for this regime. Large strain increments in the transition regime (steps 3–5) produce clear, localized strain clusters, whereas small increments in the same regime (steps 5–9) yield fields characterized by smaller and less distinct clusters (note the difference in color bar scaling for steps 3–5 vs 5–9). In the critical regime, large increments (steps 9–13, 17–19) generate persistent, well-defined shear bands, starting from step 11 when the macroscale axial strain reached about 15%. Small increments (steps 13–17), however, lead to strain fields with smaller and weaker clusters compared to those under large increments (again note color bar scaling difference for steps 13–17). The contrasting behaviors—localized clusters under large strain increments and smaller, less distinct clusters under small increments—indicate that microband formation in both regimes occurs discontinuously in time.

Figure 5 shows a cross-sectional view of the incremental volumetric strain ($\Delta\varepsilon_{vol}$) field. In the elastic, transition, and critical-state regimes with large strain increments (steps 1–5, 9–13, 17–19), the incremental volumetric strain field exhibits small clusters. When a persistent shear band—identified from the $\Delta\varepsilon_{dev}$ field in Fig. 4—emerges at step 11, volumetric strain fluctuates more markedly inside the band than outside, but without a consistent dilation or compaction trend. In the transition and critical-state regimes with small strain increments (steps 5–9, 13–17), the incremental volumetric strain field also exhibits small clusters, which are weaker than those under large strain increments (note the difference in color bar scaling for steps 1–5, 9–13, 17–19 vs 5–9, 13–17).

We further computed local non-affine motion using the $D^2_{min}$ metric, following the procedure in Ref. [26]. For a particle $i$ at macroscale strain $\gamma$, the metric measures the mean-square deviation between the actual displacements of neighboring particles and those predicted by the locally best-fitting affine transformation. In this expression, $N_i$ is the number of neighbors of particle $i$ within a spherical cutoff radius $r_c = N r_p$, where $r_p$ is the mean particle radius and $N$ is a cutoff parameter. $\mathbf{r}_i(\gamma)$ and $\mathbf{r}_j(\gamma)$ are the position vectors of particle $i$ and its neighbor $j$ from the origin at strain $\gamma$, and $\mathbf{\Gamma}$ is the local affine tensor that minimizes the total squared error.

The $D^2_{min}$ of particle $i$ from macroscale strain $\gamma$ to $\gamma + \Delta\gamma$, where $\Delta\gamma$ is the incremental macro strain, is defined as:

$$D^2_{min}(\gamma, \Delta\gamma) = \frac{1}{N_i}\sum_j^{N_i} \left\| \mathbf{r}_j(\gamma + \Delta\gamma) - \mathbf{r}_i(\gamma + \Delta\gamma) \right. \\ \left. - \mathbf{\Gamma}\left[\mathbf{r}_j(\gamma) - \mathbf{r}_i(\gamma)\right] \right\|^2 \quad (3)$$

where $\|\cdot\|$ denotes the $L_2$ norm and $\mathbf{\Gamma}$ is computed as:

$$\mathbf{\Gamma} = \mathbf{X} \cdot \mathbf{Y}^{-1} \quad (4)$$

with

$$\mathbf{X} = \sum_j^{N_i} \left[\mathbf{r}_j(\gamma + \Delta\gamma) - \mathbf{r}_i(\gamma + \Delta\gamma)\right] \otimes \left[\mathbf{r}_j(\gamma) - \mathbf{r}_i(\gamma)\right] \quad (5)$$

$$\mathbf{Y} = \sum_j^{N_i} \left[\mathbf{r}_j(\gamma) - \mathbf{r}_i(\gamma)\right] \otimes \left[\mathbf{r}_j(\gamma) - \mathbf{r}_i(\gamma)\right] \quad (6)$$

In this study, $D^2_{min}$ was computed incrementally for each pair of successive deformation steps, using a cutoff radius of $5 r_p$ to define neighbors. Unlike the approach in Ref. [26], rattler particles, which have contact numbers of two or less, were retained because the experimental dataset lacked sufficient contact information to justify their removal. The $D^2_{min}$ values, computed for individual particles, were mapped to the tetrahedral mesh, with each particle center corresponding to a mesh vertex. Each tetrahedron was then assigned the mean $D^2_{min}$ of its four vertex particles. This nodal-averaged field was then used for visualization.

Figure 6 presents a cross-sectional view of the incremental non-affine motion ($D^2_{min}$) field, which exhibits spatial patterns similar to those of the incremental deviatoric strain ($\Delta\varepsilon_{dev}$) field shown in Figure 4.

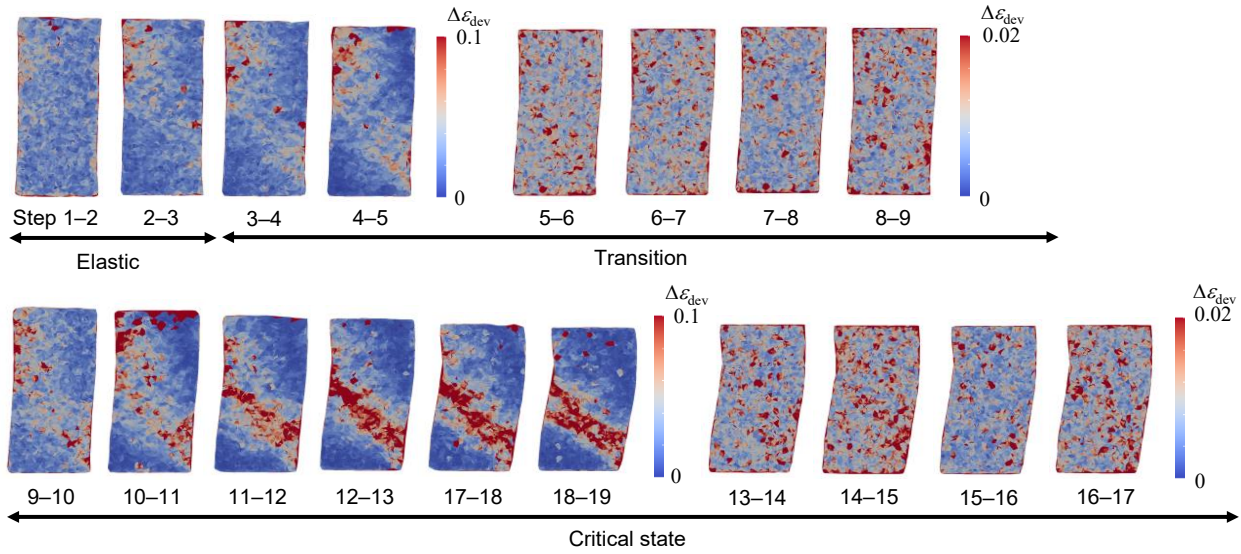

Figure 4. Cross-sectional views of the incremental deviatoric strain fields ($\Delta\varepsilon_{dev}$) across all deformation steps. Color indicates the magnitude of incremental deviatoric strain, with red representing higher values. Colorbar ranges vary to enhance contrast, particularly for steps with smaller macroscopic strain increments (e.g., steps 5–9 and 13–17).

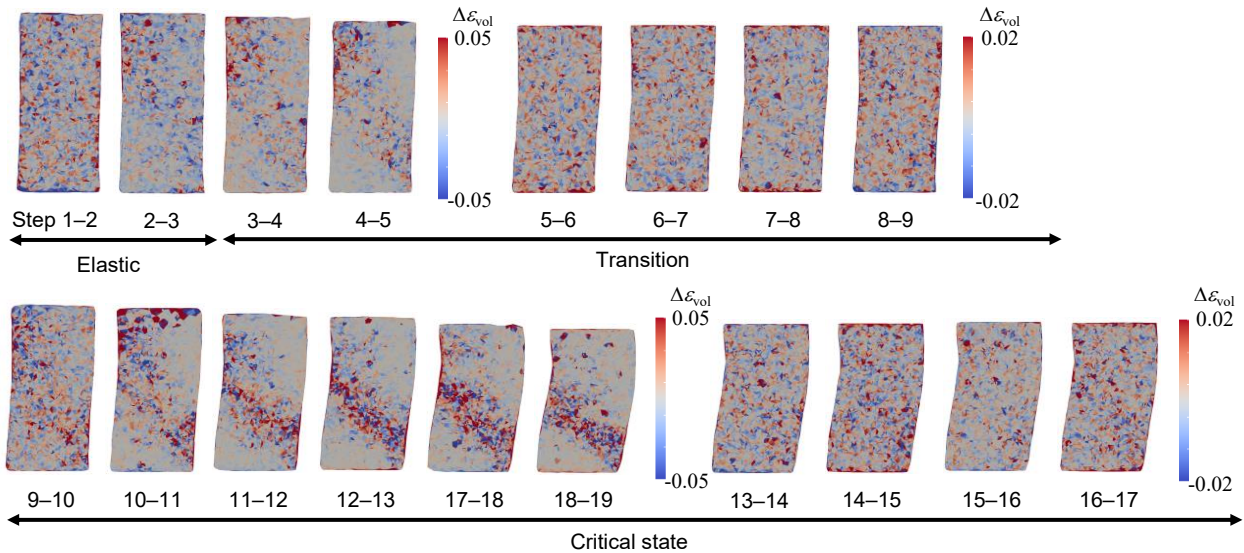

Figure 5. Cross-sectional views of the incremental volumetric strain fields ($\Delta\varepsilon_{vol}$) across all deformation steps. Color indicates the magnitude of incremental volumetric strain, with red representing higher values. Colorbar ranges vary to enhance contrast, particularly for steps with smaller macroscopic strain increments (e.g., steps 5–9 and 13–17).

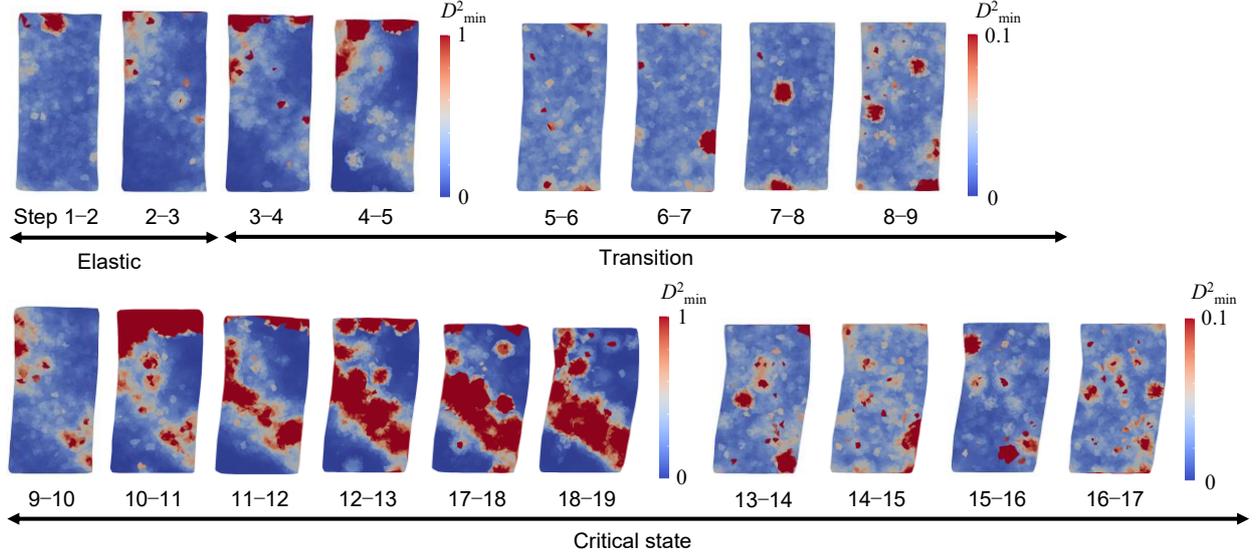

Figure 6. Cross-sectional views of the incremental non-affine motion ($D^2_{min}$) fields across all deformation steps. Color indicates the magnitude of $D^2_{min}$, with red representing higher values. Colorbar ranges vary to enhance contrast, particularly for steps with smaller macroscopic strain increments (e.g., steps 5–9 and 13–17).

3.3 Dynamic Heterogeneity

To study spatial correlations of the local non-affine motion during the experiment, we defined the following four-point spatial dynamic correlation function:

$$G_4^{DH}(r^*, t_i) = \left[ \langle D^2_{min}(\mathbf{r}, t_i, t_{i+1}) D^2_{min}(\mathbf{s}, t_i, t_{i+1}) \rangle \right.$$
$$- \langle D^2_{min}(\mathbf{r}, t_i, t_{i+1}) \rangle \langle D^2_{min}(\mathbf{s}, t_i, t_{i+1}) \rangle \right] \quad (7)$$
$$/ \left[ \text{std}(D^2_{min}(\mathbf{r}, t_i, t_{i+1})) \text{std}(D^2_{min}(\mathbf{s}, t_i, t_{i+1})) \right]$$

where $r^* = r/r_p$ is the nondimensional scalar separation distance. Here, $r$ denotes the scalar separation distance between two spatial points $\mathbf{r}$ and $\mathbf{s}$ such that $\|\mathbf{r} - \mathbf{s}\| = r$. $D^2_{min}(\mathbf{r}, t_i, t_{i+1})$ is the non-affine displacement measure of a particle located at position $\mathbf{r}$ (defined at timestep $t_i$), computed between timesteps $t_i$ and $t_{i+1}$. The operator $\langle \cdot \rangle$ denotes a spatial average over all particle pairs with separation distance $r$, and the operator $\text{std}(\cdot)$ denotes the standard deviation.

In practice, $G_4^{DH}(r^*, t_i)$ in Eq. (7) was computed using Spearman's rank correlation coefficient, which measures correlations in the relative ordering of local $D^2_{min}$ values without assuming a linear relationship. This nonparametric approach captures monotonic trends in particle-scale dynamics. To ensure sufficient statistical sampling, the correlation was evaluated over all particle pairs with separation distances in the range $\|\mathbf{r} - \mathbf{s}\|/r_p \in [r^*, r^* + \Delta r^*]$, where $\Delta r^*$ is the bin width for $r^*$, rather than at a single fixed $r^*$.

Figure 7 shows the resulting $G_4^{DH}(r^*, t_i)$ values for selected deformation steps $t_i$. In all cases, $G_4^{DH}(r^*, t_i)$ decreases with $r^*$, consistent with the spatially localized correlations associated with DH. For step 17, only particles inside the shear band were analyzed to study the characteristics of this localized deformation zone.

To identify which particles belong to the shear band in this analysis, we selected tetrahedra with $\Delta\varepsilon_{dev}$ greater than 1.5 times the volume-weighted mean of $\Delta\varepsilon_{dev}$ computed over the entire sample. The vertex positions of these tetrahedra were then used to fit a weighted regression plane, with Gaussian weights applied according to the in-plane distance from the average position of the selected vertices. Particles within $\pm 8r_p$ of this plane (i.e., total thickness of $16r_p$) were considered to reside inside the shear band while those located beyond this range were considered outside the shear band. This approach is consistent with recent methods that identify shear bands based on incremental deviatoric strain and fit a regression plane to regions of high $\Delta\varepsilon_{dev}$ [19].

Figure 8 shows the dynamic correlation length $\xi$, obtained by fitting the four-point spatial dynamic correlation function $G_4^{DH}(r^*, t_i)$, computed as described above, to the exponential form:

$$G_4^{DH}(r^*, t_i) = a \cdot \exp\left(-\frac{r^*}{\xi}\right) \quad (8)$$

A larger value of $\xi$ indicates longer-range DH, whereas a smaller $\xi$ implies more localized correlations.

Larger values of $\xi$ are observed during the transition and critical state regimes with large strain increments (steps 3–4, 9–12, and 17–18), indicating significant long-range DH. In the critical state regime, $\xi$ is greater outside the shear band than inside (e.g., steps 11–12 and 17–18), suggesting that the apparent DH is mainly driven by rigid-like motion outside the shear band. In contrast, $\xi$ remains small during the early elastic regime with large increment (step 1), and during the transition and critical state regimes with small strain increments (steps 5–8 and 13–16), indicating that DH is weak or effectively absent under these conditions. It is noteworthy that $\xi$ is large in the late elastic regime with a large strain increment (step 2)—comparable to that observed in the transition regime with large strain increments—even though the system is macroscopically elastic, suggesting that DH can emerge at the microscale prior to observable softening in macroscale behavior. The values of $\xi$ inside the shear band during large-increment steps in the critical state regime are comparable to those observed in the early elastic regime, indicating that the $D^2_{min}$ field exhibits clustering on small length scales, and that DH is minimal in the shear band, even under the macroscopic critical state.

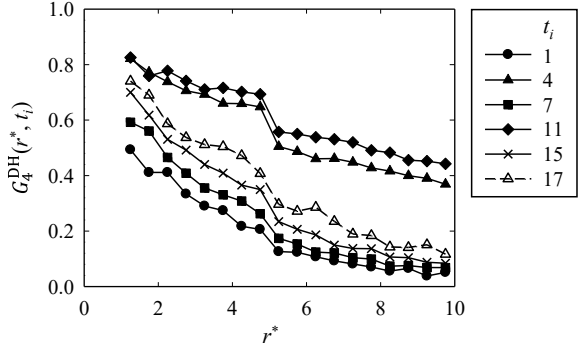

Figure 7. Four-point spatial dynamic correlation function $G_4^{DH}(r^*, t_i)$ as a function of nondimensional distance $r^*$ for selected deformation steps $t_i$. Correlations were computed based on $D^2_{min}$ values of particle pairs within distance bins $[r^*, r^* + \Delta r^*]$. For step 17, only particles inside the shear band were used.

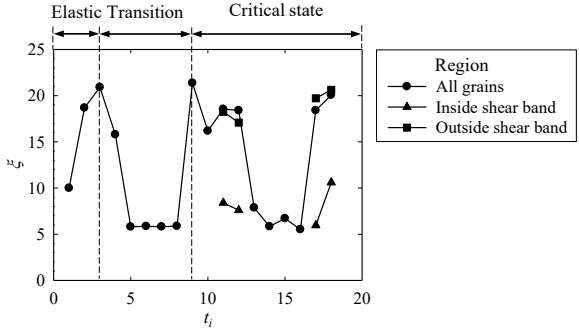

Figure 8. Dynamic correlation length $\xi$ as a function of deformation step $t_i$, obtained from exponential fits to $G_4^{DH}(r^*, t_i)$.

### 3.4 Dynamic Facilitation

Following the approach in Ref. [9], in which the mobility transfer function was defined to quantify DF in glassy systems, we computed an analogous facilitation ratio $F(r^*)$ based on $D^2_{min}$ to quantify DF in granular materials. A particle was classified as active at time $t_i$ if its non-affine displacement $D^2_{min}$ between $t_{i-1}$ and $t_i$ was within the top 10% of all particles. Newly activated particles at $t_{i+1}$ were defined as those present in both $t_i$ and $t_{i+1}$ that were inactive at $t_i$ (below the top 10% threshold for the interval $t_{i-1}$–$t_i$) but became active at $t_{i+1}$ (within the top 10% for the interval $t_i$–$t_{i+1}$). Particle identities were matched between frames using their unique IDs to ensure one-to-one correspondence.

For each active particle at time $t_i$ (referred to as a seed), we computed the radial density distribution of newly activated particles, $P_{seed}(r^*)$, as

$$P_{seed}(r^*) = \left\langle \frac{1}{\Delta r^*} N_{new}^{(i)}(r^*, r^* + \Delta r^*) \right\rangle_i \quad (9)$$

where $N_{new}^{(i)}(r^*, r^* + \Delta r^*)$ is the number of newly activated particles in the shell $[r^*, r^* + \Delta r^*]$ around the $i$-th seed, $r^* = r / r_p$ is the nondimensional scalar distance normalized by the mean particle radius $r_p$, and $\langle \cdot \rangle_i$ denotes averaging over all seeds. The distribution was normalized by the shell width $\Delta r^*$ to yield the radial density distribution.

As a baseline, we randomly sampled the same number of particles as the newly activated particles from the entire system and computed their radial density distribution relative to the same seed particles. This process was repeated 200 times using random resampling (without replacement), and the resulting distributions were averaged to obtain $P_{rand}(r^*)$.

The facilitation ratio was then defined as:

$$F(r^*) = \frac{P_{seed}(r^*)}{P_{rand}(r^*)} \quad (10)$$

where $F(r^*) > 1$ indicates that newly activated particles are more likely to be located near seeds than expected by chance.

In Figure 9, the facilitation ratio $F(r^*)$ is plotted as a function of the nondimensional distance $r^*$ for selected deformation steps $t_i$, where $t_i$ refers to the time at which the seed particles are identified (i.e., active between $t_{i-1}$ and $t_i$), and the newly activated particles are identified as active between $t_i$ and $t_{i+1}$. Our results show that DF is clearly present during the elastic and transition regimes with large strain increments (e.g., steps 2 and 4), as evidenced by elevated $F(r^*)$ values substantially above unity. In contrast, during the critical state regime under large strain increments (e.g., step 11), $F(r^*)$ remains close unity or only slightly above, indicating a pronounced weakening—or possible absence—of DF. This reduction is also observed inside the shear band (e.g., step 18), where $F(r^*)$ shows minimal deviation from one. Additionally, under small strain increments (e.g., steps 7 and 14) in the transition and critical state regimes, $F(r^*)$ remains near unity, further indicating a lack of facilitation.

These results suggest that DF is most pronounced in the early stages of deformation. Notably, even during elastic deformation (step 2), DF remains high, indicating that particle rearrangements already exhibit a non-random spatial distribution at this early stage, well before the onset of the transition regime. In contrast, in the critical state regime, the system likely reaches a dynamically stable configuration in which particle rearrangements are weakly correlated or effectively stochastic. The lack of facilitation inside the shear band further implies that plastic activity in this zone is intermittent and lacks strong spatial memory. The observed suppression of DF under small strain increments also suggests that a minimum deformation magnitude is required to activate DF.

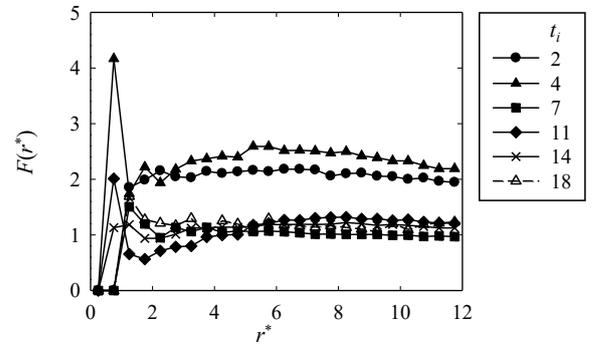

Figure 9. Facilitation ratio $F(r^*)$ as a function of nondimensional distance $r^*$ for selected deformation steps $t_i$. For step 18, only particles inside the shear band were used.

### 3.5 Temporal Correlation of Non-Affine Motion

To quantify the temporal persistence of local non-affine motion during the experiment, we defined the following four-point temporal dynamic correlation function:

$$G_4^T(t_i) = \left[ \left\langle D^2_{min}(\mathbf{r}, t_{i-1}, t_i) D^2_{min}(\mathbf{r}, t_i, t_{i+1}) \right\rangle \right.$$
$$\left. - \left\langle D^2_{min}(\mathbf{r}, t_{i-1}, t_i) \right\rangle \left\langle D^2_{min}(\mathbf{r}, t_i, t_{i+1}) \right\rangle \right] \quad (11)$$
$$/ \left[ std(D^2_{min}(\mathbf{r}, t_{i-1}, t_i)) std(D^2_{min}(\mathbf{r}, t_i, t_{i+1})) \right]$$

Here, $\langle \cdot \rangle$ denotes a global average over particles. In practice, $G_4^T(t_i)$ in Eq. (11) was computed using Spearman's rank correlation coefficient between $D^2_{min}(\mathbf{r}, t_{i-1}, t_i)$ and $D^2_{min}(\mathbf{r}, t_i,$ $t_{i+1}$), which measures monotonic relationships in local dynamics without imposing linear or model-specific assumptions.

Figure 10 shows the resulting $G_4^T(t_i)$ values as a function of deformation step $t_i$. High correlation values are observed during the transition and critical state regimes under large strain increments (steps 4, 10–12 and 18), indicating strong temporal persistence of non-affine motion. Notably, $G_4^T(t_i)$ at step 3—computed from the elastic regime (step 2–3) followed by the transition regime (step 3–4), both involving large strain increments—is also high, indicating that temporal persistence of local non-affine motion can emerge while the system is crossing from the elastic regime into the transition regime. In contrast, $G_4^T(t_i)$ at step 2 is low, likely because the microscale behaviors differ between the early elastic regime (step 1–2), where the field is spatially homogeneous, and the subsequent interval (step 2–3), where microbands begin to appear. At timesteps where the strain increment size changes (steps 5, 9, 13, and 17), the correlations are notably low, indicating that changes in strain increment size in either direction—from large to small or from small to large—suppress temporal correlations of local non-affine motion. For timesteps during small strain increments in the transition and critical state regimes (steps 6–8 and 14–16), the correlations remain relatively low, but are in general higher than those observed at steps where the increment size changes. In the critical state regime (e.g., steps 12 and 18), correlation values are generally higher outside the shear band than inside, indicating that the apparent temporal persistence is largely due to long-lived, rigid-like motion outside the shear band. Lower values inside the shear band imply that particle rearrangements are intermittent and exhibit weak temporal correlations. Notably, at step 18, when the shear band is well established, the correlation inside the shear band is lower than at step 12, when the shear band is less developed.

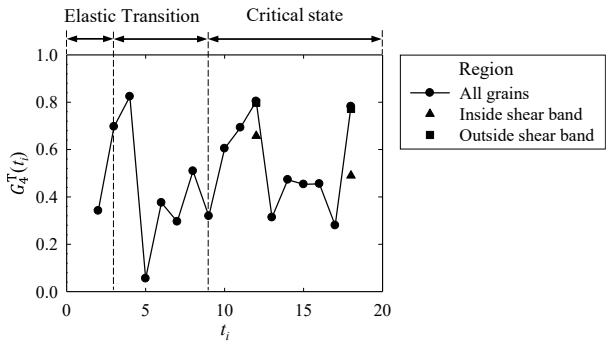

Figure 10. Four-point temporal dynamic correlation function $G_4^T(t_i)$ as a function of deformation step $t_i$.

## 4 DISCUSSION

Under large strain increments, both DH and DF clearly emerge in the transition regime, likely because microscale slip events interact, forming spatially correlated clusters of rearrangements. The persistence of DH outside the shear band during the critical state suggests that rigid-like motion outside the shear band can maintain spatial correlations even after localization. In contrast, both DH and DF inside the shear band remain weak during the critical state, as discussed below. Under small strain increments, however, both DH and DF are suppressed, indicating that a minimum deformation threshold is required to activate facilitation pathways and sustain spatial correlations. Since temporal persistence of non-affine motion is still observed at small increments, stress relaxation during imaging is unlikely to be the primary cause. Instead, the suppression appears to arise from insufficient driving, consistent with trends in glassy systems where local mobility is still present but collective heterogeneity vanishes under weak forcing [27,28].

To interpret why both DH and DF inside the shear band remain weak during the critical state, we consider the following explanations. Keys et al. [12] reported that correlation length grows with density. Candelier et al. [16] demonstrated that DF becomes less conserved and plays a diminishing role for structural relaxation near the granular glass transition. In their interpretation, the facilitation time scale is shorter than the structural relaxation time at high density, so avalanches remain local and facilitation over structural relaxation times is no longer conserved. However, in our case, we did not observe clear compaction or dilatation in the incremental volumetric strain field inside the shear band, suggesting that density change is unlikely to be the dominant factor. Thus, density driven scenarios proposed in Refs. [12,16] cannot fully account for our results. Instead, we speculate that the difference arises because microscale dynamics inside the shear band during the critical state are qualitatively different from those in the transition regime.

Alternatively, Shrivastav et al. [29] reported that shear band formation under shear is preceded by a directed percolation transition. In their framework, localized hot spots of mobility first coalesce into clusters, which at a critical strain percolate through the system. This directed percolation marks a qualitative change: the system evolves from a regime dominated by transient hot spots and micro-bands to one characterized by an emergent shear band. Thus, the establishment of a critical-state shear band can be interpreted as a phase-like transition to a shear-banded state, distinct from the transient clusters observed in the transition regime. Similarly, Karimi et al. [30] showed that the orientation of correlated plastic activity prior to shear band formation differs from that of the permanent shear band in the critical state, suggesting that microscopic clusters in the transition regime and fully developed shear bands in the critical state are governed by different microscale mechanisms. This perspective supports our interpretation that the weak DH and DF inside the shear band during the critical state reflect a qualitatively different microscale behavior compared to the transition regime.

The dynamic correlation length in the transition regime (approximately $20r^*$) is comparable to the 6–7 particle diameters reported in Ref. [11] and the 5–10 molecular diameters reported for glassy materials in Ref. [6]. This similarity indicates that the spatial scale of correlated dynamics in granular materials is on the same order as in glassy systems, underscoring the universality of DH across disordered matter. A more complete mechanical interpretation of these trends would require direct force-chain and stress measurements at the grain scale, which we consider an interesting avenue for future work beyond the present kinematic analysis.

## 5 CONCLUSIONS

We examined DH and DF in 3D sheared granular materials using *in-situ* X-ray tomography. Under large strain increments, both DH and DF were pronounced during the transition regime, with DH persisting into the critical state mainly outside the shear band. Under small strain increments, both were suppressed. These findings demonstrate that the presence of correlated dynamics depends strongly on strain increment and deformation regime.

Beyond these specific observations, our study extends previous 2D granular analyses to 3D systems and applies DH and DF—concepts widely used in glassy systems—to granular materials, highlighting the broader relevance of spatiotemporal correlations across different disordered systems.

## 6 ACKNOWLEDGEMENTS

This research was supported by the U.S. National Science Foundation CAREER Grant No. CBET-1942096 and a Johns Hopkins University Catalyst Award.

## 7 DATA AVAILABILITY

The data that support the findings of this article are available in Ref. [31].

## 8 REFERENCES

[1] Le, L. A., Nguyen, G. D., Bui, H. H., & Andrade, J. E. (2022). Localised failure of geomaterials: How to extract localisation band behaviour from macro test data. *Géotechnique*, 72(7), 596–609.

[2] Falk, M. L., & Langer, J. S. (1998). Dynamics of viscoplastic deformation in amorphous solids. *Physical Review E*, 57(6), 7192.

[3] Berthier, L. (2011). Dynamic heterogeneity in amorphous materials. *Physics*, 4, 42.

[4] Kegel, W. K., & van Blaaderen, A. A. (2000). Direct observation of dynamical heterogeneities in colloidal hard-sphere suspensions. *Science*, 287(5451), 290–293.

[5] Weeks, E. R., Crocker, J. C., Levitt, A. C., Schofield, A., & Weitz, D. A. (2000). Three-dimensional direct imaging of structural relaxation near the colloidal glass transition. *Science*, 287(5453), 627–631.

[6] Ediger, M. D. (2000). Spatially heterogeneous dynamics in supercooled liquids. *Annual Review of Physical Chemistry*, 51(1), 99–128.

[7] Zhang, P., Maldonis, J. J., Liu, Z., Schroers, J., & Voyles, P. M. (2018). Spatially heterogeneous dynamics in a metallic glass forming liquid imaged by electron correlation microscopy. *Nature Communications*, 9(1), 1–7.

[8] Wang, X., Xu, W. S., Zhang, H., & Douglas, J. F. (2019). Universal nature of dynamic heterogeneity in glass-forming liquids: A comparative study of metallic and polymeric glass-forming liquids. *The Journal of Chemical Physics*, 151(18).

[9] Elmatad, Y. S., & Keys, A. S. (2012). Manifestations of dynamical facilitation in glassy materials. *Physical Review E*, 85(6), 061502.

[10] Herrero, C., & Berthier, L. (2024). Direct numerical analysis of dynamic facilitation in glass-forming liquids. *Physical Review Letters*, 132(25), 258201.

[11] Dauchot, O., Marty, G., & Biroli, G. (2005). Dynamical heterogeneity close to the jamming transition in a sheared granular material. *Physical Review Letters*, 95(26), 265701.

[12] Keys, A. S., Abate, A. R., Glotzer, S. C., & Durian, D. J. (2007). Measurement of growing dynamical length scales and prediction of the jamming transition in a granular material. *Nature Physics*, 3(4), 260–264.

[13] Le Bouil, A., Amon, A., McNamara, S., & Crassous, J. (2014). Emergence of cooperativity in plasticity of soft glassy materials. *Physical Review Letters*, 112(24), 246001.

[14] Viallon-Galinier, L., Combe, G., Richefeu, V., & Picardi Faria Atman, A. (2018). Emergence of shear bands in confined granular systems: Singularity of the q-statistics. *Entropy*, 20(11), 862.

[15] Candelier, R., Dauchot, O., & Biroli, G. (2009). Building blocks of dynamical heterogeneities in dense granular media. *Physical review letters*, 102(8), 088001.

[16] Candelier, R., Dauchot, O., & Biroli, G. (2010). Dynamical facilitation decreases when approaching the granular glass transition. *Europhysics Letters*, 92(2), 24003.

[17] Houdoux, D., Amon, A., Marsan, D., Weiss, J., & Crassous, J. (2021). Micro-slips in an experimental granular shear band replicate the spatiotemporal characteristics of natural earthquakes. *Communications Earth & Environment*, 2(1), 90.

[18] Shahin, G., & Hurley, R. C. (2022a). Micromechanics and strain localization in sand in the ductile regime. *Journal of Geophysical Research: Solid Earth*, 127(11), e2022JB024983.

[19] Hurley, R. C., Shahin, G., Kuwik, B. S., & Lee, K. (2023). Assessing continuum plasticity postulates with grain stress and local strain measurements in triaxially compressed sand. *Proceedings of the National Academy of Sciences*, 120(32), e2301607120.

[20] Phan, Q. T., Bui, H. H., Nguyen, G. D., & Nicot, F. (2024). Strain localization in the standard triaxial tests of granular materials: Insights into meso-and macro-scale behaviours. *International Journal for Numerical and Analytical Methods in Geomechanics*, 48(5), 1345–1371.

[21] Thakur, M. M., Henningsson, N. A., Engqvist, J., Autran, P. O., Wright, J. P., & Hurley, R. C. (2023). On mesoscale modeling of concrete: Role of heterogeneities on local stresses, strains, and representative volume element. *Cement and Concrete Research*, 163, 107031.

[22] Wadell, H. (1935). Volume, shape, and roundness of quartz particles. *The Journal of Geology*, 43(3), 250–280.

[23] Shahin, G., & Hurley, R. C. (2022b). HP-TACO: A high-pressure triaxial compression apparatus for in situ x-ray measurements in geomaterials. *Review of Scientific Instruments*, 93(11), 113902.

[24] Stamati, O., Andò, E., Roubin, E., Cailletaud, R., Wiebicke, M., Pinzon, G., Couture, C., Hurley, R. C., Caulk, R., Caillerie, D., et al. (2020). SPAM: Software for practical analysis of materials. *Journal of Open Source Software*, 5(51), 2286.

[25] Zhang, B., & Regueiro, R. A. (2015). On large deformation granular strain measures for generating stress–strain relations based upon three-dimensional discrete element simulations. *International Journal of Solids and Structures*, 66, 151–170.

[26] Lee, K., & Hurley, R. C. (2025). Interplay between forces, particle rearrangements, and macroscopic stress fluctuations in sheared 2D granular media. *Physical Review E*, 112(1), 015413.

[27] Priezjev, N. V. (2013). Heterogeneous relaxation dynamics in amorphous materials under cyclic loading. *Physical Review E*, 87(5), 052302.

[28] Priezjev, N. V. (2014). Dynamical heterogeneity in periodically deformed polymer glasses. *Physical Review E*, 89(1), 012601.

[29] Shrivastav, G. P., Chaudhuri, P., & Horbach, J. (2016). Yielding of glass under shear: A directed percolation transition precedes shear-band formation. *Physical Review E*, 94(4), 042605.

[30] Karimi, K., & Barrat, J. L. (2018). Correlation and shear bands in a plastically deformed granular medium. *Scientific reports*, 8(1), 4021.

[31] Lee, K., Kuwik, B, & Hurley, R. C. (2025). Dataset for article titled "Dynamic Heterogeneity and Facilitation in Sheared Granular Materials: Insights from 3D Triaxial Testing" [Dataset], Zenodo (2025), doi: 10.5281/zenodo.17187358.